\documentclass{article}
\pdfoutput=1 
\usepackage{spconf,amsmath,graphicx,bm,amssymb}
\usepackage{amsmath,amsfonts,xcolor,graphicx,hyperref,bm, adjustbox, tabularx}
\usepackage{hyperref}
\usepackage{xcolor}
\usepackage[font=small,labelfont=bf]{caption}
\usepackage[square,sort&compress,comma,numbers]{natbib}
\usepackage{multirow}
\usepackage{float}
\usepackage{dblfloatfix}
\usepackage[]{color-edits}
\addauthor{ma}{red}
\addauthor{by}{blue}

\title{Improved Supervised Training of Physics-Guided Deep Learning Image Reconstruction with Multi-Masking}
 \name{\begin{tabular}{c}Burhaneddin Yaman$^{\star \dagger}$, Seyed Amir Hossein Hosseini$^{\star \dagger}$, Steen Moeller$^{\dagger}$ and Mehmet Ak\c{c}akaya$^{\star \dagger}$
 \end{tabular}}

 \address{$^{\star}$ Electrical and Computer Engineering, University of Minnesota, Minneapolis, MN, USA \\
      $^{\dagger}$ Center for Magnetic Resonance Research, University of Minnesota, Minneapolis, MN, USA\\
            }

\begin{document}
\maketitle
\begin{abstract}
 Physics-guided deep learning (PG-DL) via algorithm unrolling has received significant interest for improved image reconstruction, including MRI applications. These methods unroll an iterative optimization algorithm into a series of regularizer and data consistency units. The unrolled networks are typically trained end-to-end using a supervised approach. 
 Current supervised PG-DL approaches use all of the available sub-sampled measurements in their data consistency units. Thus, the network learns to fit the rest of the measurements. In this study, we propose to improve the performance and robustness of supervised training by utilizing randomness by retrospectively selecting only a subset of all the available measurements for data consistency units. The process is repeated multiple times using different random masks during training for further enhancement. Results on knee MRI show that the proposed multi-mask supervised PG-DL enhances reconstruction performance compared to conventional supervised PG-DL approaches. 
\end{abstract}


\section{Introduction}
\label{sec:intro}
Lenghty acquisition times remains a challenge in MRI often causing patient discomfort or artifacts in the reconstruction. Hence, accelerating MRI remains an ongoing research topic. MRI reconstruction is based on an inverse problem that incorporates the physics of the data acquisition via the encoding operator. Direct recovery from acquired sub-sampled measurements is often ill-posed, and thus regularizers are often used for solving this inverse problem. Such regularized reconstruction is conventionally solved using iterative algorithms, such as gradient descent (GD), or proximal gradient descent (PGD) and variable splitting methods that alternate between data consistency (DC) and a proximal operator for the regularizer \cite{fessler_SPM}.

Recently, deep learning has emerged as an alternative approach for solving such inverse problems in MRI \cite{DongLiang, RAKI, Hammernik,Hemant,JongChulYeeDLMagPhase,MortezaLoss,Knoll_SPM, yaman_SSDU_MRM,Amir_EMBC,Kamilov_Rare,Sabuncu_SamplingPattern,LeslieYing_SPM}. Among deep learning methods, physics-guided deep learning (PG-DL) methods, also known as algorithm unrolling \cite{monga2019algorithm}, have gained interest, as it incorporates the forward model for acquired measurements to the network architecture. PG-DL unrolls iterative optimization algorithms consisting of DC and regularizer units for a fixed number of iterations \cite{Hammernik,deepADMMnet,Hemant,QinUnrollnetwork,Hosseini_JSTSP,deepcomplexmri}.
The conventional fixed sparsifying regularizers in these algorithms are replaced with neural networks, while gradient descent or conjugate gradient (CG) methods are employed in DC units \cite{Hammernik,Hemant}. These network are then typically trained end-to-end in a supervised manner using a ground-truth reference. While PG-DL approaches may differ based on the optimization algorithm they unroll, all existing approaches use all of the available sub-sampled measurements in their DC units. Thus, the unrolled network learns to fit the rest of the unacquired measurements while the DC units ensure consistency with the acquired ones.

In this study, we propose to improve the performance and robustness of supervised training of PG-DL methods by using a multi-masking operation on the available measurements by retrospectively selecting a random subset of measurements from the original sub-sampled data for use in DC units multiple times. DC units in the unrolled network only use these subset of the measurements as opposed to conventional supervised training, where all of them are used. We hypothesize that including such random masking may inherently improve the robustness of the trained PG-DL algorithm, which will learn to fit to larger sets of different measurements for different masks, reducing residual aliasing artifacts, while not requiring any other modifications to the training process. Results on knee MRI show that the proposed multi-mask supervised PG-DL approach enhances the reconstruction performance  compared to conventional supervised PG-DL by further removing residual artifacts, and improving quantitative metrics, such as SSIM and PSNR.

\section{Methods} 
\label{sec:meterialsandmethods}
\subsection{Algorithm Unrolling for MRI Reconstruction}
Let $\mathbf{y}_\Omega$ be the acquired k-space measurements with $\Omega$ denoting the undersampling pattern, and ${\bf x}$ be the image to recover. The forward model for acquired measurements is given as 
\begin{equation}\label{first_eq}
	{\bf y}_{\Omega} = {\bf E}_{\Omega} {\bf x} + {\bf n},
\end{equation}
where ${\bf E}_{\Omega}: {\mathbb C}^{M} \to {\mathbb C}^P$ is the  encoding operator containing the partial Fourier sampling, coil sensitivities and the undersampling pattern, and ${\bf n} \in {\mathbb C}^P$ is the measurement noise. Recovery of $\mathbf{x}$ from $\mathbf{y}_\Omega$ is formulated as
\begin{equation}\label{Eq:recons1}
\arg \min_{\bf x} \|\mathbf{y}_{\Omega}-\mathbf{E}_{\Omega}\mathbf{x}\|^2_2 + \cal{R}(\mathbf{x}),
\end{equation}
where the first term enforces data consistency, and $\cal{R}(\mathbf{\cdot})$ is a regularizer. Optimization techniques \cite{fessler_SPM} such as variable splitting with quadratic penalty \cite{Hemant, QinUnrollnetwork, Yaman_MultiMask_SSDU} can be employed to cast Eq. (\ref{Eq:recons1}) into an alternating minimization problem as 
\begin{subequations}
\begin{align}
& \mathbf{z}^{(i-1)} = \arg \min_{\bf z}\mu \lVert\mathbf{x}^{(i-1)}-\mathbf{z}\rVert_{2}^2 +\cal{R}(\mathbf{z})\label{Eq:recons3a}
\\
& \mathbf{x}^{(i)} = \arg \min_{\bf x}\|\mathbf{y}_{\Omega}-\mathbf{E}_{\Omega}\mathbf{x}\|^2_2 +\mu\lVert\mathbf{x}-\mathbf{z}^{(i-1)}\rVert_{2}^2\label{Eq:recons3b}
\end{align}
\end{subequations}
where $\mathbf{z}^{(i)}$ is an auxiliary intermediate variable and $\mathbf{x}^{(i)}$ is the desired image at iteration $i$. In PG-DL methods, this iterative algorithm is unrolled for a fixed number of iterations \cite{monga2019algorithm}. Sub-problem Eq. (\ref{Eq:recons3a}) is solved implicitly with a neural network, while the DC unit in sub-problem (\ref{Eq:recons3b}) is solved using conventional linear methods \cite{Hammernik, QinUnrollnetwork, Hemant}.

\subsection{Supervised PG-DL Training}
Supervised PG-DL algorithms aims to map network input (sub-sampled k-space/distorted image) to  ground-truth reference (fully-sampled k-space/artifact free image) by training neural networks on a database of acquired slices. Let ${\bf y}_{\textrm{ref}}^i$ represents the fully-sampled data for subject $i$ and $f({\bf y}_{\Omega}^i, {\bf E}_{\Omega}^i; {\bm \theta})$ be the the unrolled network output for input sub-sampled k-space data ${\bf y}_{\Omega}^i$, in which the network is parameterized by ${\bm \theta}$. The objective function for supervised PG-DL training in k-space is formulated as
\begin{equation} \label{loss_eq}
    \min_{\bm \theta} \frac1N \sum_{i=1}^{N} \mathcal{L}( {\bf y}_{\textrm{ref}}^i, \:{\bf E}_{\textrm{full}}^i f({\bf y}_{\Omega}^i, {\bf E}_{\Omega}^i; {\bm \theta})),
\end{equation}
where $N$ is the number of slices in the database, ${\bf E}_{\textrm{full}}^i$ is the multi-coil encoding operator that transform network output to k-space, ${\bf y}_{\textrm{ref}}^i$ is the fully-sampled ground-truth k-space and $\mathcal{L}(\cdot, \cdot)$ denotes the loss function. Learnt parameters for the unrolled network during training are subsequently used to reconstruct unseen undersampled test data.

\begin{figure}[t]
  	\begin{minipage}[b]{1.0\linewidth}
  		\centering
  		\centerline{\includegraphics[width=8.5 cm]{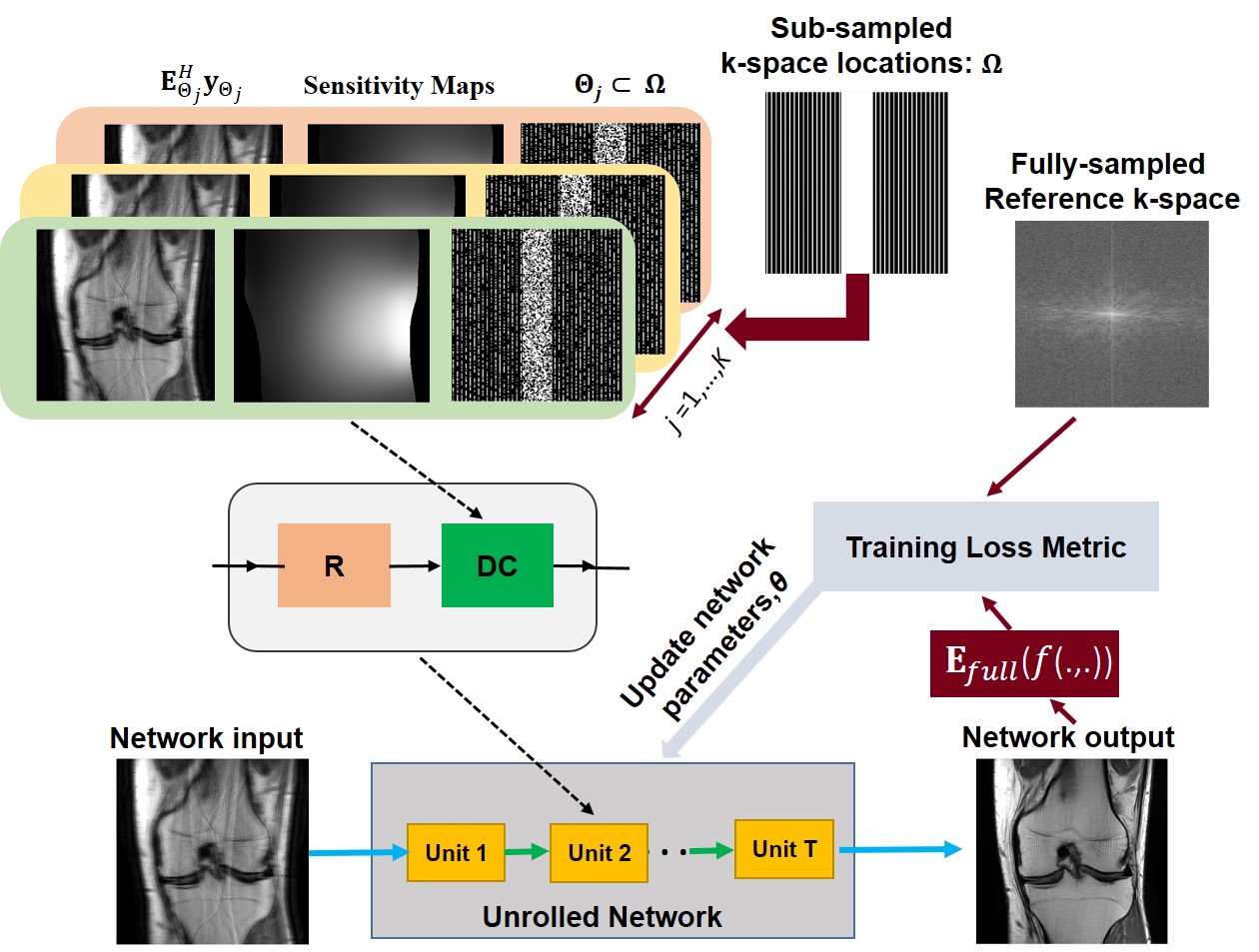}}
  	\end{minipage}

  	\caption{The proposed multi-mask supervised PG-DL MRI reconstruction. The sub-sampled acquired measurements for each scan, $\Omega$, are retrospectively further masked into subsets $\Theta_j \subset \Omega$, $j \in \{1, \dots, K\}$, multiple times for use in DC units. Loss is performed in k-space by comparing the fully-sampled reference k-space with the corresponding reconstructed k-space of the network output. Network parameters are subsequently updated based on the training loss.}
  	\label{fig:MultiMask_Architecture}
  	\vspace{-.3cm}
\end{figure}
\subsection{Proposed Multi-Mask Supervised PG-DL Training}
With supervised training as in Eq. (\ref{loss_eq}), PG-DL approaches learn to fit the unacquired measurements during training, while their DC units ensure consistency with all the acquired sub-sampled measurements. Thus, we hypothesize that the quality and robustness of reconstruction can be further improved by masking the entries available to the DC units during supervised training \cite{Yaman_MultiMask_SSDU}. To this end, instead of using all available sub-sampled measurements $\Omega$ during data consistency, we propose to use a random subset of available measurements in the DC units by retrospectively masking $\Omega$ multiple times as illustrated in Figure \ref{fig:MultiMask_Architecture}. Formally, the acquired sub-sampled measurements for each slice in the dataset is retrospectively masked $K$ times as
\begin{equation}
     \Theta_j \subset \Omega , \ \ j \in \{1,\dots, K\}.
\end{equation}
Hence, the loss function in Eq. (\ref{loss_eq}) is reformulated as 
\begin{equation}
    \min_{\bm \theta} \frac{1}{N\cdot K} \sum_{i=1}^{N}\sum_{j=1}^{K} \mathcal{L}\Big({\bf y}_{\textrm{ref}}^i, \: {\bf E}_{\textrm{full}}^i \big(f({\bf y}_{\Theta_j}^i, {\bf E}_{\Theta_j}^i; {\bm \theta}) \big) \Big).
\end{equation}

 \begin{figure*}[!t]
 \begin{center}
           \includegraphics[trim={0 0 0 0},clip, width=7 in]{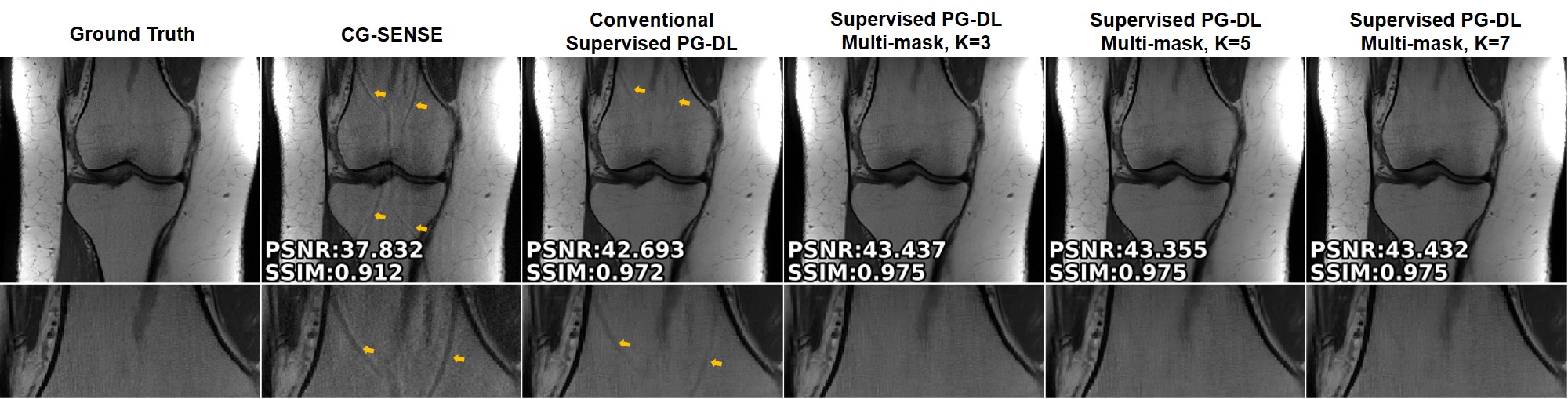}
     \end{center}
      \vspace{-.35cm}
  	\caption{Reconstruction results at R = 4 with uniform undersampling using CG-SENSE, proposed multi-mask and conventional supervised PG-DL approaches. CG-SENSE suffers from significant noise amplification and residual artifacts shown with yellow arrows. Conventional supervised PG-DL approach also exhibits residual artifacts for this slice. Proposed multi-mask supervised PG-DL approach outperforms the conventional PG-DL approach by succesfully removing the residual artifacts, while $K$=3 masks achieve the highest metrics.
  	}
  	\label{fig:supervised_multimask}
\end{figure*}
\subsection{Network Training Details}
\label{sec:24}
The proposed multi-mask supervised and conventional supervised PG-DL approaches are implemented by unrolling iterative sub-problems in (\ref{Eq:recons3a})-(\ref{Eq:recons3b}) for 10 iterations. Each iteration contains DC and regularizer units, which are respectively implemented with CG \cite{Hemant} and the ResNet structure used in \cite{yaman_SSDU_MRM}. Coil sensitivity maps are generated from central 24$\times$24 ACS using ESPIRiT \cite{ESPIRIT}. Training is performed by using Adam optimizer with a learning rate of $5\cdot 10^{-4}$. Network is trained to minimize a normalized $\ell_1$ - $\ell_2$ loss over 100 epochs with a batch size of 1 \cite{yaman_SSDU_MRM}. All experiments for PG-DL approaches are performed using Tensorflow in Python.

\subsection{Imaging Experiments}
\label{ssec:invivodatasets}
 Fully-sampled coronal proton density weighted knee MRI dataset were obtained from the New York University (NYU) fastMRI initiative database \cite{fastmri}. Imaging was performed on a clinical 3T system (Magnetom Skyra; Siemens, Erlangen, Germany) with a 15-channel coil-array using 2D turbo spin-echo sequences. Relevant imaging parameters were \cite{Hammernik}:
matrix size = $320 \times 368$, in-plane resolution = $0.49 \times 0.44 \textrm{ mm}^2$, slice thickness = $3$ mm. Training was performed on 300 slices from 15 subjects using 20 central slices from each. Testing was performed on all slices of 10 different subjects, leading to total of 380 slices. 

Fully-sampled raw data was retrospectively subsampled to rate, R = 4 using a uniform undersampling pattern with 24 ACS lines. For the proposed method, $\Theta_j$ of each partition was selected based on a uniformly random distribution. Furthermore, $|\Theta_j| / |\Omega|$, where $|\cdot|$ denotes the cardinality of the index set, was chosen as 0.6 based on a previous self-supervised learning study \cite{Yaman_MultiMask_SSDU}. The number of partitions for each slice was empirically investigated for $K \in \{3,5,7\}$. Proposed multi-mask supervised PG-DL approach was compared with conventional supervised PG-DL and conjugate gradient SENSE (CG-SENSE) \cite{cgsense}. Testing was performed both for uniform and random undersampling patterns. PSNR and SSIM were used for quantitative evaluation on the test dataset.
 
  \begin{figure*}[!t]
 \begin{center}
           \includegraphics[trim={0 0 0 0},clip, width=7 in]{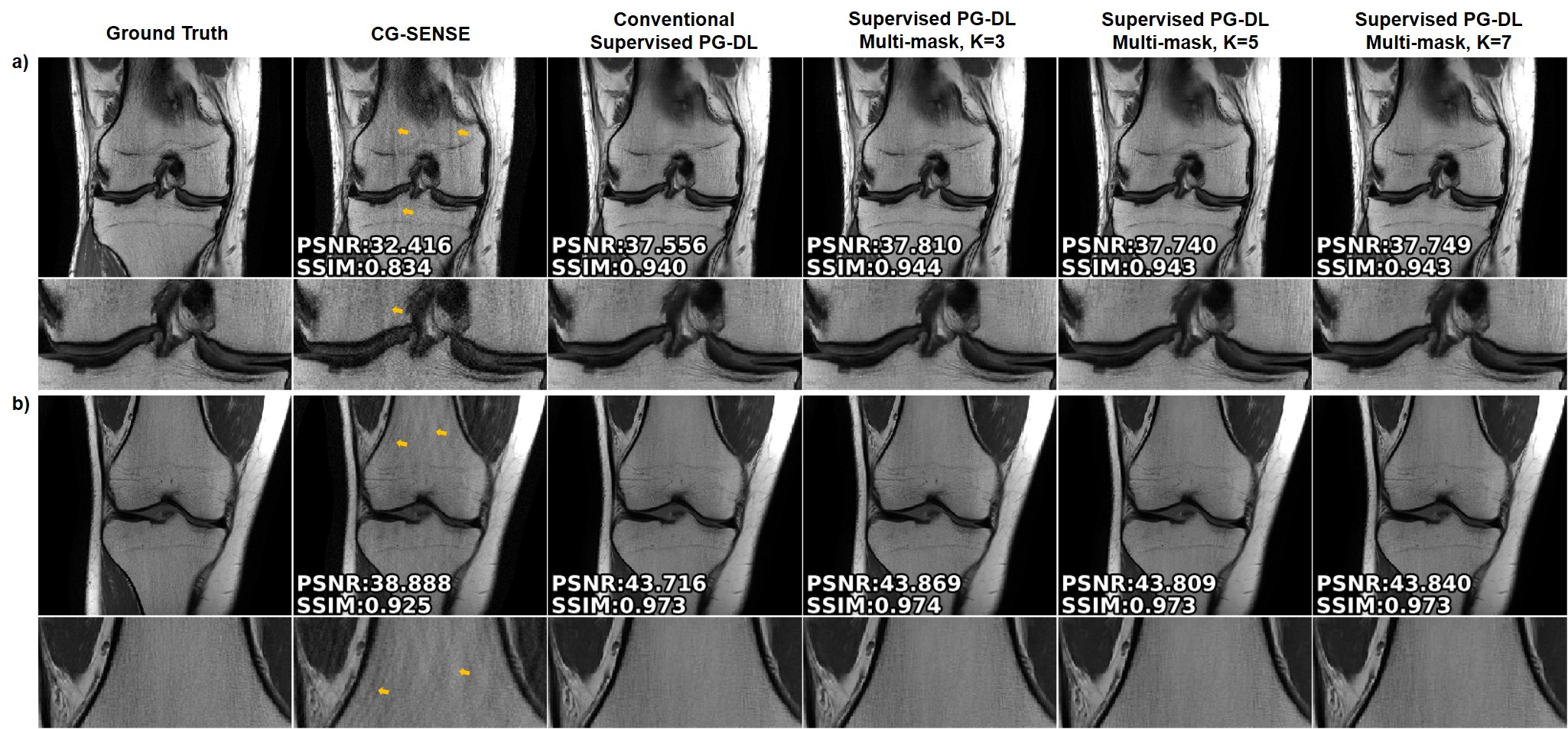}
     \end{center}
      \vspace{-.35cm}
  	\caption{Reconstruction results for two different slices at R = 4 using CG-SENSE, proposed multi-mask and conventional supervised PG-DL approaches for a) uniform and b) random undersampling scenarios. For both sampling patterns, CG-SENSE suffers from noise amplification and residual artifacts shown with yellow arrows, while conventional supervised and proposed multi-mask supervised PG-DL approaches perform closely without depicting any residual artifacts. For both random and uniform undersampling, proposed multi-mask supervised at $K$=3 achieves the highest PSNR and SSIM.
  	}
  	\label{fig:supervised_multimask_random}
\end{figure*}

 \vspace{-.3cm}
\begin{table*}[!b]
\begin{adjustbox}{width=0.9\textwidth,center}
\begin{tabular}{|l|l|l|l|l|l|l|}
\hline
& \begin{tabular}[c]{@{}l@{}}Quantitative \\ $\:\:$ Metric\end{tabular} & \ \ \ \ \ \ \ \ \ CG-SENSE                    & \begin{tabular}[c]{@{}l@{}}\ \ \ \ \ \  Conventional\\ \ \ \ Supervised PG-DL\end{tabular} & \begin{tabular}[c]{@{}l@{}}\ \ \ \  Supervised PG-DL\\ \ \ \ \ Multi-mask, K=3\end{tabular} & \begin{tabular}[c]{@{}l@{}}\ \ \ \ Supervised PG-DL\\ \ \ \ \ Multi-mask, K=5\end{tabular} & \begin{tabular}[c]{@{}l@{}}\ \ \ \ Supervised PG-DL\\ \ \ \ \  Multi-mask, K=7\end{tabular} \\ \hline
\multirow{2}{*}{\begin{tabular}[c]{@{}l@{}}Uniform\\ Sampling\end{tabular}} & \ \ \ \ SSIM                                                            & 0.881 {[}0.849, 0.919{]}    & 0.965 {[}0.955, 0.977{]}                                                & \bf{0.968 {[}0.957, 0.979{]}}                                                  & 0.967 {[}0.956, 0.977{]}                                                   & 0.966 {[}0.955, 0.977{]}                                                   \\ \cline{2-7} 
            & \ \ \ \ PSNR                                                            & 35.329 {[}33.771, 37.413{]} & 40.822 {[}39.308, 42.290{]}                                             & \bf{41.199 {[}39.417,42.702{]}}                                                & 41.031 {[}39.372, 42.727{]}                                                & 41.069 {[}39.341, 42.634{]}                                                \\ \hline
\multirow{2}{*}{\begin{tabular}[c]{@{}l@{}}Random\\ Sampling\end{tabular}}  & \ \ \ \ SSIM                                                            & 0.901 {[}0.874, 0.930{]}    & 0.968 {[}0.955, 0.978{]}                                                & \bf{0.970 {[}0.958, 0.979{]} }                                                 & 0.969 {[}0.957, 0.978{]}                                                   & 0.968 {[}0.957, 0.978{]}                                                   \\ \cline{2-7} 
            & \ \ \ \  PSNR                                                            & 36.122 {[}34.672, 38.135{]} & 40.987 {[}39.435, 43.068{]}                                             &\bf{41.216 {[}39.620, 43.315{]}}                                               & 41.099 {[}39.529, 43.338{]}                                                & 41.087 {[}39.534, 43.298{]}                                                \\ \hline
\end{tabular}
\end{adjustbox}
\vspace{-.3cm}

  \caption{The median and interquartile range (25$^\textrm{th}$-75$^\textrm{th}$ percentile) of SSIM and PSNR metrics on test dataset for uniform and random undersampling cases. For both sampling scenarios, CG-SENSE is outperformed by PG-DL approaches in terms of both SSIM and PSNR. The proposed multi-mask supervised PG-DL achieves improved metrics compared to conventional supervised PG-DL approach for all $K>1$. For multi-mask supervised PG-DL, $K=3$ leads to the highest PSNR and SSIM values for both uniform and random undersampling scenarios. 
  }
  \label{tbl:ssim_psnr_table}
\end{table*}

\section{Results}
Figure \ref{fig:supervised_multimask} shows reconstruction results from a representative test slice using  conventional supervised PG-DL, i.e. $K = 1$, and the proposed multi-mask supervised PG-DL for $K \in \{3, 5, 7\}$, as well as conventional CG-SENSE. CG-SENSE suffers from significant residual artifacts and noise amplification.  Conventional supervised PG-DL approach also displays residual artifacts on this slice, marked with yellow arrows.  Proposed multi-mask supervised PG-DL achieves a better reconstruction performance by further suppressing these residual artifacts for all choices of $K$. 

Figure \ref{fig:supervised_multimask_random}a and \ref{fig:supervised_multimask_random}b display reconstruction results on two different subjects using uniform and random undersampling, respectively. CG-SENSE suffers from residual artifacts in both cases, with random undersampling showing fewer artifacts, due to the incoherent nature of aliasing. For both of these slices, proposed multi-mask supervised PG-DL and conventional supervised PG-DL do not suffer from any visible artifacts, while closely performing with each other.

Table \ref{tbl:ssim_psnr_table} summarizes the median and interquartile ranges (25$^\textrm{th}$-75$^\textrm{th}$ percentile) of SSIM and PSNR values of all reconstruction methods for uniform and random undersampling patterns. For both uniform and random undersampling scenarios, CG-SENSE performs worse than PG-DL approaches, while the proposed multi-mask supervised PG-DL approaches achieves improved metrics compared to conventional supervised PG-DL approach. Proposed multi-mask supervised PG-DL achieves the best metrics for $K=3$, which is consistent with the observations in Figures \ref{fig:supervised_multimask} and \ref{fig:supervised_multimask_random}.

\section{Discussion and Conclusion}
\label{sec:discussion}
In this study, we proposed a multi-mask supervised PG-DL approach, which retrospectively masks the undersampled measurements in the DC units multiple times to enhance the reconstruction quality. Results on knee MRI showed that the proposed method successfully removes residual artifacts that may occur in some test datasets using conventional supervised PG-DL. Moreover, it preserves the image quality and further improves the quantitative metrics of conventional supervised PG-DL if there are no artifacts in the reconstruction. 

Supervised PG-DL provides improved reconstruction quality compared to conventional clinical approaches \cite{Hammernik,Hemant,Hosseini_JSTSP,LeslieYing_SPM}. However, it may still exhibit residual artifacts on some test slices. Hence, we hypothesized that bringing a degree of randomness into supervised learning can further improve its robustness and reconstruction performance. The proposed multi-masking approach achieves improved robustness, as the additional randomization in the DC units enforce trained networks to learn different sets of measurements. Importantly, this is achieved without requiring any major change in the training process.

The proposed multi-mask supervised PG-DL was shown to suppress residual artifacts that were observed in some test slices using conventional supervised PG-DL. Additionally, for all $K$ values, there was also a quantitative improvement in metrics compared to conventional supervised PG-DL. These improvements come only at the expense of $K$-fold increased training time, which does not have any impact on the testing time. While proposed multi-masking ($K>$1) improves performance over conventional supervised PG-DL with $K$=3 achieving the best results quantitatively, increasing $K$ further does not necessarily lead to more improvements. These observations on selection of $K$ are consistent with the literature on data augmentation, where the optimal size of the post-augmented data remains an open research problem \cite{shorten2019survey}.
\section{acknowledgements}
\label{sec:acknowledgments}
This work was partially supported by NIH R01HL153146,
P41EB027061, U01EB025144; NSF CAREER CCF-1651825. Knee MRI data were obtained from the NYU fastMRI initiative database \cite{fastmri}. NYU fastMRI database was acquired with the relevant institutional review board approvals as detailed in \cite{fastmri}. NYU fastMRI investigators provided data but did not participate in analysis or writing of this report. A listing of NYU fastMRI investigators, subject to updates, can be found at \url{fastmri.med.nyu.edu}. We also thank an anonymous reviewer of \cite{Yaman_MultiMask_SSDU} for their comments encouraging us to pursue this direction further.

\bibliographystyle{IEEEbib}
\bibliography{reference}
\end{document}